# RF surface resistance tuning of superconducting niobium via thermal diffusion of native oxide

E.M. Lechner,[1,a)] J.W. Angle,[2)] F.A. Stevie,[3)], M.J. Kelley,[1,2)] C.E. Reece[1)] and A.D. Palczewski,[1)]

[1]*Thomas Jefferson National Accelerator Facility, Newport News, Virginia 23606, USA*

[2]*Virginia Polytechnic Institute and State University, Blacksburg, Virginia 24061, USA*

[3]*Analytical Instrumentation Facility, North Carolina State University, Raleigh, North Carolina 27695, USA*

[a)] Author to whom correspondence should be addressed: lechner@jlab.org

Recently, Nb superconducting radio frequency cavities vacuum heat treated between 300-400°C for a few hours have exhibited very high quality factors (~$5\times10^{10}$ at 2.0 K). New secondary ion mass spectrometry measurements of O, N and C show this enhancement in RF surface conductivity is primarily associated with interstitial O alloying via dissolution and diffusion of the native oxide. We use a theory of oxide decomposition and O diffusion to quantify previously unknown parameters crucial in modeling this process. RF measurements of a vacuum heat treated Nb superconducting radio frequency cavity confirm the minimized surface resistance (higher $Q_0$) previously expected only from 800°C diffusive alloying with N.

Superconducting radio frequency (SRF) technology is and will be an enabling technology for current and future major particle accelerators used to facilitate fundamental scientific inquiries [1-10]. Its employment is due to extreme efficiency and ability to operate in high duty cycle or continuous wave modes [11]. Efforts to increase the efficiency of Nb SRF resonant cavities via interstitial alloying with nitrogen [12-15], titanium [16,17], or other impurities [18] have yielded a significant reduction of the surface resistance, $R_s$, by a factor of 2–4. Performance enhancement in interstitially alloyed cavities is inherently a complex multifaceted problem. In this space, the performance is dependent on the interplay between electron mean free path [19-24], enhanced sensitivity to trapped flux [19,25,26], a normal-conducting hydride precipitate blocking effect [27-30], and density of states tuning [31-36]. With interstitial alloying, these factors have often worked in concert to produce resonant cavities with unprecedented low surface resistance (high quality factor), but often with lower maximum supportable field amplitude.

Recently, Ito [37] investigated the effect of vacuum heat treatment of Nb superconducting radio-frequency cavities between 200 °C and 800 °C. In the range of 300 °C to 400 °C, the cavities exhibited pronounced decrease of $R_s$ with field, extremely high quality factors, and reduced quench fields, all typical qualities previously associated with nitrogen-alloyed Nb cavities. Qualitatively, these effects were assumed to be due to oxygen diffusion from the native Nb oxide. Ito's work is strikingly similar to the work done at Fermi National Accelerator Laboratory (FNAL), where cavities were vacuum heat-treated, in some cases exposed to nitrogen, and RF tested without exposure to atmosphere [38,39]. Initial time-of-flight secondary ion mass spectrometry (TOF-SIMS) measurements by FNAL observed $Nb_2O_5$ dissolution and a qualitative increase in nitrogen concentration near the surface ~10 nm deep. From these observations it was assumed that nitrogen was the primary alloying diffusant. It was noted explicitly in [38] that their conclusions on the quality factor enhancement were in disagreement with the oxygen dissolution and migration results of previous works.

Early works in vacuum heat treating Nb SRF cavities in the temperature range of 250 °C to 300 °C showed that the native oxide layer dissolved and a reduced BCS surface resistance, $R_{BCS}$, was measured, likely due to tuning the electron mean free path by oxide dissolution and oxygen diffusion [40,41]. In these studies, the residual resistance increased, washing out any benefit of $R_{BCS}$ reduction. Later, experiments showed that baking cavities around 100 °C improved performance [42,43]. An oxygen dissolution and diffusion model introduced by Ciovati [44] to describe the migration of oxygen during the 120 °C bake was developed [45]. However, some dissolution and diffusion parameters in that model were too unconstrained for predictive O concentration diffusion during higher temperature vacuum heat treatments. We present new secondary ion mass spectrometry measurements on vacuum heat treated Nb small samples and quantify the O, N, and C impurity concentration depth profiles. The O concentration profiles are analyzed using the model of Ciovati to quantify previously unknown parameters related to oxide dissolution into the bulk.

Samples were prepared following a similar process to the one outlined by Ito [37]. The samples were cut from Tokyo Denkai ASTM 6 Nb stock procured using the XFEL/007 specification [46]. The stock was first vacuum annealed at 900°C to promote grain growth following the same procedure as the 1.3 GHz single-cell cavity, SC-16, used for RF validation below [47]. First, each sample was nano-polished (NP) by a vendor to a surface roughness, $R_a$, of ~2 nm to provide sufficiently flat samples for SIMS measurements. Post NP, the samples received a 600°C/10hr heat-treatment to remove bulk hydrogen caused by the mechanical polishing before a 20 µm electropolish with the typical $HF/H_2SO_4$ solution at 13°C [48]. During each heat treatment, the samples were housed in a double-walled Nb foil container to minimize any furnace contamination [49]. Samples were baked for various times and temperatures, shown in table I, to explore the parameter space of the oxide dissolution and oxygen diffusion processes.

Table I. Vacuum heat treatment temperature and duration for samples examined by SIMS.

| Sample | Temperature (°C) | Time (hr) |
|---|---|---|
| NL409 | 300 | 2.3 |
| NL411 | 350 | 2.7 |
| NL431 | 140 | 48.3 |
| NL438 | 280 | 3.0 |
| NL439 | 240 | 0.9 |
| NL440 | 300 | 2.6 |
| NL447 | 140 | 12.3 |
| NL448 | 330 | 0.5 |
| NL449 | 220 | 20 |

SIMS measurements were made using a CAMECA 7f Geo magnetic sector SIMS instrument. The primary ion beam is comprised of $Cs^+$ using an accelerating potential of 5 kV and sample potential of -3 kV for an impact energy of 8 keV. This ion beam is rastered over an area of 150 μm × 150 μm and the data collected from a 63 μm × 63 μm area in the center of the larger raster. Proper quantitation of SIMS depth profiles requires the use of implant standards in order to convert the ion signal to impurity concentration [50]. Here we used implant standards to quantify the O, C and N composition of RF penetration layer and beyond by detecting $^{16}O^-$, $^{12}C^-$ and $^{107}(NbN)^-$ secondary ions in conjunction with a $^{93}Nb^-$ reference signal. The implant standards used in these SIMS experiments were dosed with C, N and O at $2\times10^{15}$ atoms/cm$^2$ at 135 keV, 160 keV and 180 keV, respectively by Leonard Kroko Inc. SIMS depth profiles were acquired to the background O, N and C levels for all samples.

The impurities most likely to be alloyed into Nb during a vacuum thermal diffusion process are O, N and C, however, in the temperature range probed here, N and C have diffusion coefficients at least 2–3 orders of magnitude less than O [51,52]. Figure 1a shows the calibrated SIMS depth profiles of O, N and C in a sample vacuum heat treated with the SRF cavity SC-16 described later in the text at 300 °C for 2.6 hours. We observe small concentrations of C and N, but a large increase in O concentration from the expected background O concentration of ~0.004 O at. % seen in similar samples [53], in agreement with our measurements of a freshly electropolished sample shown in figure 1b. In the absence of a gaseous oxygen source, this high concentration of subsurface interstitial oxygen is due to oxygen dissolution from the 3-6 nm thick native pentoxide [54-56]. Depth profiles up to ~13 μm were made to quantify the strength of the oxygen source and parameters related to oxide dissolution into the bulk. We analyzed our SIMS-derived concentration depth profiles using the model of oxide dissolution and oxygen diffusion in Nb calculated analytically by Ciovati [44]. This approach considers an initial interstitial oxygen concentration at the surface and a finite source term due to oxide dissolution.

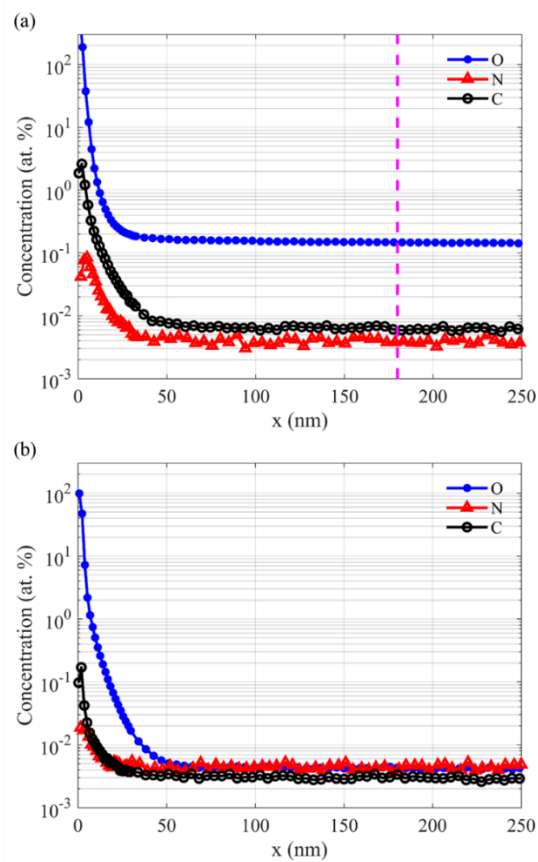

FIG. 1. (a) Typical calibrated SIMS depth profile of a sample vacuum heat treated with cavity SC-16 described later in the text. This sample was vacuum heat treated at 300 °C for 2.6 hours presenting a large enhancement in oxygen with minor amounts of carbon and nitrogen at the near surface. The vertical dashed magenta line represents the electropolishing depth that cavity SC-16 underwent to exclude any ingress of nitrogen or carbon after vacuum heat treatment. (b) Typical calibrated SIMS depth profile of an electropolished sample with no subsequent heat treatment.

The SIMS depth profiles and theoretical fits are shown in figure 2. Least-squares fits to the SIMS depth profile were performed using shared parameters across all SIMS depth profiles, and the background O concentration was taken from the end of the depth profile. These measurements provide insight into the oxygen dissolution parameters that have been previously unknown.

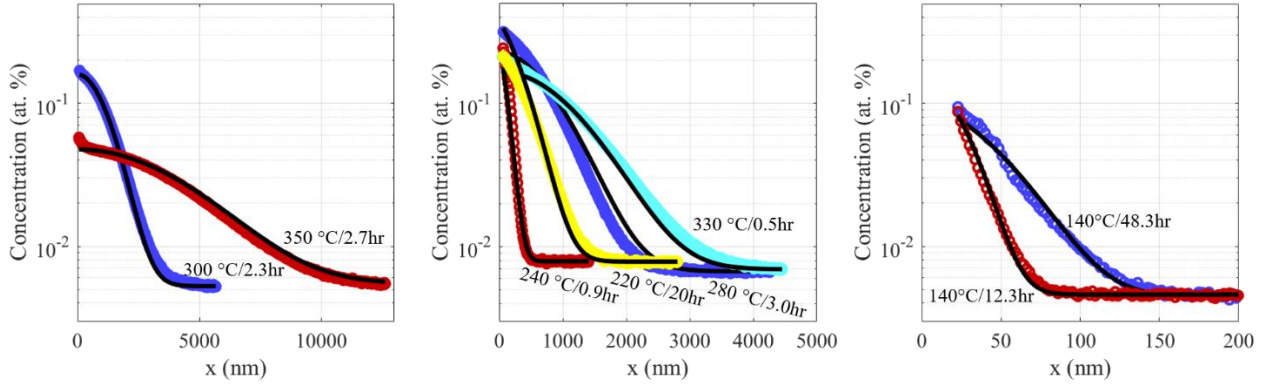

FIG. 2. Calibrated SIMS O depth profiles of samples vacuum heat treated at various temperatures and times in circles and the fit of Ciovati's model in black. The same model parameters described in the text are used for all theoretical fits.

To review, Ciovati's model accounts for an oxide dissolution from a finite source and diffusion into a semi-infinite slab of Nb which is described by the one-dimensional diffusion equation with a source term

$$\frac{\partial c(x,t,T)}{\partial t} = D(T)\frac{\partial^2 c(x,t)}{\partial x^2} + q(x,t,T). \qquad (1)$$

Equation (1) can be separated into solutions for the initial interstitial oxygen concentration $v(x,t)$ and oxygen introduced into the system due to oxide dissolution $u(x,t)$ such that $c(x,t) = v(x,t) + u(x,t)$. Those solutions to this system are [44]

$$v(x,t,T) = \frac{v_0}{\sqrt{\pi D(T)t}} e^{-x^2/(4D(T)t)} + c_\infty \qquad (2)$$

$$u(x,t,T) = \frac{u_0}{\sqrt{\pi D(T)}} \int_0^t \frac{k(T)e^{-k(T)s}}{\sqrt{t-s}} e^{-x^2/(4D(T)(t-s))} ds. \qquad (3)$$

Where $v_0$ is the initial interstitial concentration near the surface and $u_0$ quantifies the oxygen contribution from the oxide layer dissolution. Both the oxide dissolution rate constant and the diffusion coefficient are assumed to have Arrhenius behavior such that $D(T) = D_0 e^{-E_{aD}/RT}$ and $k(T) = Ae^{-E_{ak}/RT}$. $c_\infty$ is the background bulk concentration of interstitial oxygen, not included in Ciovati's model. From the least-squares fit to all O concentration depth profiles, $v_0$ was found to be $3.5 \pm 0.1$ O at. % nm, approximately a factor of 3 less than the value of 10 O at. % nm estimated by Ciovati. $E_{aD}$ was found to be 119.9 ± 0.3 kJ/mol which is larger than typically reported values for oxygen diffusion in Nb at higher temperatures [44,51,57]. This increase in $E_{aD}$ may be due to interactions between interstitial oxygen atoms which have been observed in internal friction measurements [52,58,59]. We measured $D_0$ to be 0.075 ± 0.005 cm$^2$/s which is in agreement with a similarly reported diffusion activation energy [60]. The reported frequency factor, $A$, and activation energy, $E_{ak}$, for oxygen dissolution varies substantially. Some reports have estimated the oxygen dissolution activation energy as low as 58 kJ/mol or 71 kJ/mol [61,62] and others have estimated it much higher at ~135 kJ/mol [63,64]. The surface probes used in those studies cannot reliably determine the activation energy and frequency factor of oxide dissolution into the metal, because the amount of oxygen entering the surface is on the order of 0.1%, often below any meaningful detection limit with only a few nm probe depth. We find an activation energy of 131 ± 3 kJ/mol for oxide dissolution into the bulk in agreement with the pentoxide dissolution activation energy measured and implemented by others [44,63,64]. The frequency factor 0.9 ± 0.6 ×10$^9$ 1/s is also in approximate agreement with the value reported by Ciovati [44] and close to the upper limit of niobium pentoxide dissolution estimated by King [64]. The extreme sensitivity of dynamic SIMS is required to determine the activation energy and frequency factor of oxygen dissolution. The oxygen contribution from the oxide layer dissolution, $u_0$, was measured to be 200 ± 2 O at. % nm which provides the first experimental determination of this parameter and is significantly different from its initial estimation of 1000 O at. % nm [44].

To confirm the effect of interstitial oxygen alloying in Nb, an SRF cavity was vacuum heat treated at 300 °C for 2.6 hours. The temperature ramp rate was approximately 3 °C/min with a total pressure better than 1.1×10$^{-6}$ Torr during heating. The cavity cooled under vacuum to room temperature before venting. During the first attempt to RF test SC-16, contamination was suspected after an improper assembly. A 1 hour nitric soak was performed to remove potential contaminants without removing the native oxide [65]. This was followed by an electropolish removal 180 ± 20 nm at 7 °C by standard process [66,67] for the purpose of excluding any contributions to the impurity profile due to shallow migration of other interstitials like C or N to clearly remove their contribution to any alloying behavior without significantly affecting the oxygen content. We have also vacuum heat treated two other SRF cavities, to be reported on elsewhere, at temperatures of ~300 °C and successfully tested them without additional chemical treatments in agreement with other results [37-39,68]. RF measurements before and after vacuum annealing are shown in figure 3. Before interstitial O alloying, the cavity was quench limited at $B_{peak}$ of 170 mT (41 MV/m) with a minimum $R_s$ of 13 nΩ (maximum $Q_0$ of 2.2×10$^{10}$) [69]. After O alloying the cavity and electropolishing removal of 180 nm from the surface to confidently exclude any near surface adventitious nitrogen uptake, the RF test at 2 K showed a reduction of $R_s$ (pronounced $Q_0$ rise) with a minimum $R_s$ of 5.8 nΩ (maximum $Q_0$ of $4.7 \times 10^{10}$) at 70 mT (16 MV/m) and a quench field of 90 mT (21 MV/m). The field reduction of $R_s$ with the removal of 180 nm via electropolishing demonstrates that the surface resistance reduction effect here is due primarily, if not exclusively, to oxygen diffusion into the surface. It should be noted that the observed oxygen concentration in our witness sample is on the same order as the impurity concentration required to produce high quality factors in N-alloyed cavities [22,50,70]. These high quality factors have been often ascribed to mean free path tuning and hydride blocking effects for which considerable impurity concentrations are required [22,71-74]. The effect of electron mean free path tuning via impurity alloying and its effect on surface resistance has been established [19,22], is microscopically justified [20,21] and general to any impurity that modifies the electron mean free path [21,73]. Estimates of electron mean free path tuning show that O, C and N have similar effects [22,73,74]. The hydride blocking effects of O should be similar to N as well [30].

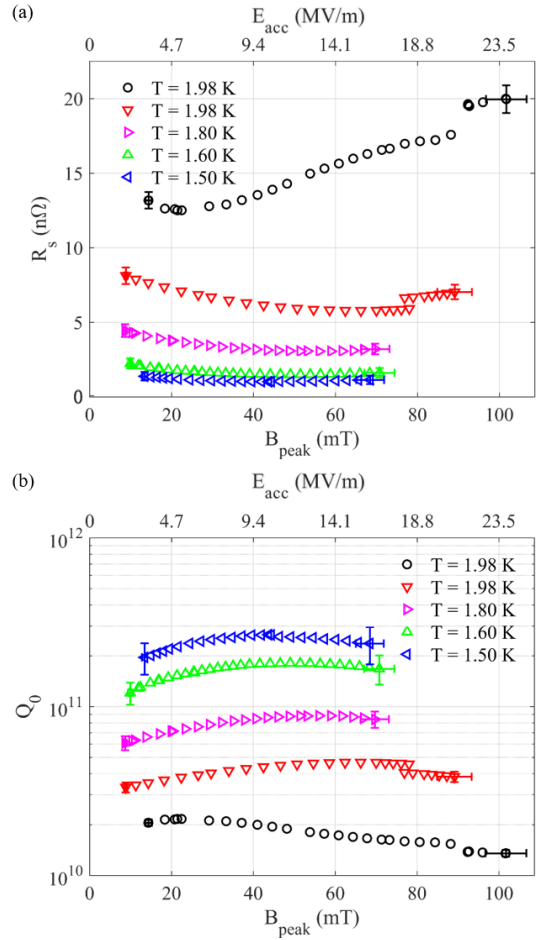

FIG. 3. (a) $R_s(B_{peak})$ and (b) $Q_0(B_{peak})$ measurements for the cavity baked at 120 °C for 12 hours in black and after vacuum furnace heat treatment at 300 °C for 2.6 hours in blue, magenta, green and red at various temperatures.

The modeling shown above, along with calibrated SIMS results and RF cavity measurements, show the intermediate-temperature (300–400 °C) vacuum heat treatment processes of Ito, FNAL and IHEP [38,39,68] may be fully explainable in terms of oxygen as the alloying diffusant. A calibrated SIMS investigation of corresponding process samples would provide a better understanding of the impurity concentrations necessary to achieve those low surface resistances. Perhaps, the observed small interstitial nitrogen content [38,39] may be a secondary effect over the 5-10 nm near surface. Interestingly, the role of adventitious oxygen in RF surfaces exposed to nitrogen at modest (200–400°C) temperatures (sometimes termed "nitrogen infusion") may have been underestimated. Near surface oxygen is observed to be nearly two orders of magnitude greater than bulk background [75-79]. Future analysis will seek to clarify if indeed oxygen is the dominant agent in such circumstances.

In conclusion, we have for the first time quantified the parameters for native niobium oxide dissolution into the niobium bulk in this temperature regime and linked this to the minimization of RF surface resistance in superconducting niobium. The experimental method outlined here could prove useful for determining parameters relevant for other native oxide dissolution processes for various metals. The native oxide dissolution and diffusion process characterized in this work, whose effects on RF performance have been examined by others [37-41,68] and confirmed here, could be an enabling break from the high temperatures required for interstitial nitrogen "doping". The oxide dissolution and diffusion model opens many avenues for surface modification of niobium and tuning of superconducting properties. The reduced furnace cleanliness required, the lack of the need to remove surface-precipitated nitrides, and the reduced need for post heat-treatment chemistry [37] will reduce the stringent requirements necessary and increase repeatability across vacuum furnace facilities. The robust modeling outlined with this paper, based on Ciovati's finite oxygen source model, promises to enable precise electron mean free path tuning from the dirty limit to clean limit by selection of vacuum heat treatment time and temperature. This simple process opens up conformal interstitial O alloying to niobium cavities of all geometries and Nb thin film resonators [80]. This process may be modified to create tailored impurity profiles [81] by performing a complex temperature ramp, multiple dissolutions, or tuning the oxygen source via simple preanodization [82]. It may also facilitate precise explorations into the role of interstitial O concentration on the afforementioned interplay of the electron mean free path, enhanced sensitivity to trapped flux, hydride blocking effects and density of states tuning to untangle their contributions to RF performance. Beyond accelerator SRF applications, such a process may be crucial to emerging technologies like devices for quantum information systems [83] as well as other applications where the surface interstitials need tuning.

## ACKNOWLEDGEMENTS

This work was coauthored by Jefferson Science Associates LLC under U.S. DOE Contract No. DE-AC05-06OR23177. This material is based on work supported by the U.S. Department of Energy, Office of Nuclear Physics in the Office of Science. The authors are grateful for support from the Office of High Energy Physics, U.S. Department of Energy under Grant No. DE-SC-0014475 to Virginia Tech for support of J. Angle.

## DATA AVAILABILITY

The data that support the findings of this study are available from the corresponding author upon reasonable request.